\documentstyle[12pt, epsf]{article}
\newcommand{\beq}{\begin{equation}}
\newcommand{\beqr}{\begin{eqnarray}}
\newcommand{\eeqr}{\end{eqnarray}}

\newcommand{\eeq}{\end{equation}}

\newcommand{\intom}{\int_{-\infty}^{\infty}\frac{d\omega}{2\pi}}

\newcommand{\sid}{\mbox{$\psi^{\dagger}$}}
\newcommand{\intK}{\int_{- \pi}^{\pi} \frac{dK}{2\pi}}
\newcommand{\intk}{\int_{- \Lambda}^{\Lambda} \frac{dk}{2\pi}}

\newcommand{\sib}{\mbox{$\overline{\psi}$}}

\begin{document}
\setlength{\baselineskip}{0.375in}
\begin{center}
\Large\bf Luttinger revisted-the renormalization group approach.
\\ \normalsize \vspace{.5in} R.Shankar \\ Sloane Laboratory of
Physics \\ Yale University\\ New Haven CT 06520 \vspace{.5in}\\
August 27,2000\\ \Large\bf Abstract
\end{center}
\normalsize \setlength{\baselineskip}{.3in}

Luttinger's contributions abound in different parts of many-body
physics. Here I review the ones that appear when one uses the
Renormalization Group (RG) to study the subject: the Luttinger
Liquid, Luttinger's Theorem (on the volume of the Fermi surface)
and the Kohn-Luttinger Theorem on the superconducting instability
of all metals as one approaches absolute zero.
\newpage
\tableofcontents \setlength{\baselineskip}{.375in}
\section{Introduction}
As someone who entered condensed matter physics at a ripe old age,
my exposure to it has been rather unconventional, guided more by
my research interests than by a standard curriculum. In this
process I noticed that the name Luttinger kept cropping up all
over the place. Here I discuss the instances
 that arose  in my
attempts to apply the Renormalization Group (RG) to understanding
many-body physics. What follows is a rather personal view of these
topics, as compared to an objective review. Given the context,
this seems reasonable.

It is amusing that the RG leads to many of these results since
Luttinger's methods were quite different and could be summarized
as follows:\begin{itemize}
\item Master field theoretic methods as applied to
many-body physics.
\item Attack the problem frontally and try to demolish it.
\end{itemize}
The RG approach is roughly as follows: \begin{itemize}
\item Set up the problem as a multiple (path) integral.
\item Keep chipping away at the integrals, saving the most
difficult (singular) part for last, with no intention of doing it.
\end{itemize}

I met Luttinger only once, following a colloqium I gave on these
topics. It was a brief encounter of the first kind, i.e., I did
not get time to find out in depth his reaction to these
alternative approaches. My own view is this. While it is  very
satisfying  to be able to recast the results of Luttinger or
Landau in RG language (which makes them easy for some of us to
 understand and extend), each time that happens, my respect for
the original inventor of these ideas only increases, for I ask:
"How did he do it without any of this machinery to help him?" I am
therefore honored to  contribute these personal remarks in a
volume dedicated to this remarkable physicist.

\section{The RG approach} About a decade ago, when I got into
the business, high $T_c$ materials had been discovered and
Anderson had challenged the community to find an alternative to
Landau's Fermi liquid theory. In one dimension, there was a
concrete case of a non-Fermi liquid, which had been dubbbed the
Luttinger liquid by Haldane \cite{Haldane}. I will return to this
topic shortly, but must add that the model was posed by
Luttinger\cite{Luttinger1} and correctly solved by  Mattis and
Lieb \cite{Mattis}, and is isomorphic to the massless Thirring
model in quantum field theory. The question in high $T_c$ was
whether two dimensional fermions behaved more like this case or
like the three dimensional case, where Fermi liquids ruled.

At that time I had already become a great fan of the RG, having
seen its awesome power in the realm of critical phenomena. I
wanted to  apply it to the two-dimensional Fermi system to
determine it fate. The basic idea was simple. First the free
system in the low energy region, {\em which for fermions is near
the Fermi surface}, would be cast in the form of a functional
integral, and an RG that left its action fixed would be
determined. Then, all interactions would be viewed as
perturbations of this fixed point and classified as relevant,
irrelevant and marginal. Only the relevant and marginal ones could
possibly  destabilize the Fermi liquid. I published my preliminary
findings in 1991 in an issue of Physica devoted to Michael
Fisher's sixtieth birthday \cite{RG1} (since I felt this was
exactly the kind of stuff Michael would enjoy) and followed it
with a very detailed RMP article in 1994 \cite{RG2}. I became
aware that Benfatto and Gallvotti \cite{Benfatto} and Feldman and
Trubowitz \cite{Feldman} had published results on this topic.
Their approach was a lot more formal than I was accustomed to.
(After the above mentioned work of Mattis and Lieb I had developed
a healthy respect for doing things carefully and realized that
sometimes the correct {\em physics} emerges only when one
understands the mathematics, in this example, unitarily
inequivalent representations, correctly. In the present case
however, I am confident  my approach has the right physics.)
Polchinski, \cite{Polchinski} who was interested in effective
field theories for particle physics, independently arrived at the
central idea in a somewhat more schematic description in 1992.
Anderson had mentioned this approach in his book\cite{PWA}.

I decided I would start with $d=1$ as a warm up to see how well
the method did. Let us begin with the fact that in $d=1$, free
spinless fermions hopping on a lattice are described by the
following hamiltonian in momentum spaces: \beqr H_0 &=& \intK \sid
(K) \psi (K) E( K)
\\ E(K) &=& -\cos K \label{freefer} \eeqr where \beq \{ \psi (K) ,
\sid (K') \} = 2 \pi \delta (K- K'). \eeq  The Fermi sea is
obtained by filling all negative energy states, i.e., those with
$|K| \le K_F = \pi /2 $, which corresponds to half-filling, the
case I will specialize on here. The Fermi surface  consists of
just two points $|K| = \pm \pi /2$. It is clear that the ground
state is a perfect conductor since  we can move a particle just
below the Fermi surface  to just above it at arbitrarily small
energy cost.

Now we   argue that at weak coupling, only modes near $\pm K_F$
will be activated. Thus we will linearize the dispersion relation
$E(K) = -\cos K$ near  these points and work with a cut-off
$\Lambda$: \beq H_0 = \sum_i \intk \psi^{\dag}_{i} (k) \psi_i (k)
k \label{freefercont} \eeq where \beqr k &=& |K| - K_F \\ i   &=&
L, R \ \ \ (left \ \ or \ \ right). \eeqr

 Next  we will write down a $T=0$ partition function for the noninteracting
fermions. This will be a Grassmann integral only over the degrees
of freedom within a cut-off $\Lambda$  of  the Fermi surface. We
will then find an   RG transformation   that lowers the cut-off
but  leaves the free-field action, $S_0$,  invariant.  With the
RG  well defined,  we will  look at the generic perturbations of
this fixed point and classify them as usual.
 If no relevant operators show up, we will still have a scale-invariant gapless
system. If, on the other hand, there are generic relevant
perturbations, we will have to  see to which new fixed point the
system  flows. (The new one could also be gapless.) The stability
analysis can be done perturbatively. {\em In particular,  if a
relevant perturbation takes us away from the original fixed point,
nothing at higher orders  can ever bring us back to this fixed
point. }

Let us then  begin with the partition function for our  system of
fermions: \beqr Z_0 &=& \int \prod_{i=L,R}\prod _{|k|<\Lambda}
d\psi_{i} (\omega k) d\sib_{i} (\omega k)e^{S_0} \\
  S_0        &=& \sum_{i = L\ R} \intk \intom \sib_{i} (\omega  k) (i\omega -
k)\psi_{i} (\omega k)\label{freeaction}
  \eeqr

 This is just a product of functional integrals for the Fermi oscillators at each
momentum with frequency $\Omega_0 (k)= k$.

  The first step in the   RG    transformation   is to integrate out all $\psi
(k\omega )$ and $\sib (k\omega )$ with
  \beq
  \Lambda / s \le |k| \le \Lambda  \label{newcut}
\eeq and {\em all }$\omega$. Thus our phase space has the shape of
a rectangle, infinite in the $\omega$ direction, but  finite in
the $k$ direction. (Consult Figure 1 for  details.) This shape
will be preserved under the RG transformation. Since there is no
real relativistic invariance here, we will make no attempt to
treat $\omega$ and $k$ on an equal footing. Allowing $\omega$ to
take all values  allows us to extract  an effective  hamiltonian
operator at any stage in the RG since  locality in time is
assured.

Since the integral is gaussian, the result of integrating out fast
modes  is just a numerical prefactor which we throw out. The
surviving modes now have their momenta going from $-\Lambda  /s $
to $\Lambda  /s$.  To make this action a fixed point  we define
rescaled variables: \beqr k' &=& sk \nonumber \\ \omega '  &=&
s\omega \nonumber \\ \psi_{i} '(k'\omega ') &=& s^{-3/2} \psi_{i}
(k\omega  )\label{rescale} \eeqr Ignoring a constant that comes
from rewriting the measure in terms of the new fields, we see that
$S_0$ is invariant under the mode elimination and rescaling
operations.

We can now consider the effect of perturbations on this fixed
point. Rather than turn on the perturbation corresponding to any
particular
 interaction (say nearest neighbor),
we will perform a more general analysis. The result for the
particular cases will be subsumed by this analysis.
\subsection{Quadratic perturbations}
First consider perturbations which are quadratic in the fields.
These must necessarily be of the form \beq \delta S_2 = \sum_{i =
L,\ R} \intk \intom \mu (k\omega  )\sib_{i} (\omega k) \psi_{i}
(\omega k) \eeq
 assuming  symmetry between left and right fermi points.

 Since this action separates into slow and fast pieces, the effect of mode
elimination is simply to reduce $\Lambda$ to $\Lambda /s$ in the
integral above. Rescaling moments and fields, we find
 \beq
 \mu' (\omega' , k', i) = s \mu (\omega , k ,i).
 \eeq
 We get this factor $s$  as a result of combining a factor $s^{-2}$ from
rewriting the  old  momenta and frequencies in terms of the  new
and a factor $s^3$ which comes from rewriting the old fields in
terms of the new.

  Let us expand $\mu$ in a Taylor series
 \beq
 \mu (k, \omega) = \mu_{00} + \mu_{10} k + \mu_{01} i\omega + \cdots +
 \mu_{nm} k^{n} (i\omega )^{m} + \cdots \label{muexp}
 \eeq
 The constant piece is a relevant perturbation:
 \beq
 \mu_{00} \longrightarrow  s\mu_{00} .
 \eeq
  This relevant flow reflects the readjustment of the Fermi sea
  to a change in chemical potential.  As for the next
two terms, they are marginal.  When we consider quartic
interactions, it will be seen that mode elimination will produce
relevant and marginal terms of the above form even if they were
not there to begin with just as $\phi^4$ theory. The way to deal
with all relevant quadratic terms will be discussed
  in a moment. The marginal terms will modify the Fermi
   velocity and rescale the field.
  As for higher order terms in  Eqn.(\ref{muexp}), they are
irrelevant under the   RG  mentioned above.

Note that in order to define the RG, we need to know the location
of the Fermi momentum $K_F$, since we zero-in on the Fermi surface
(a pair of points in this case) as the RG acts. Here again we find
a very useful result due to Luttinger: the  Fermi momentum  is
determined by the number density and the relation between them is
the same as in the free case. This result can be exploited here if
we work with fixed number of particles rather than a fixed
chemical potential. This means that we must determine the
requisite $\mu$ perturbatively as we go along. In my long article
I spell out the details of this process, but the main point is
that in the RG language, this is the way  to fine-tune  (the
coefficient of) a relevant operator to zero. This operator does
not produce a gap, but instead moves the Fermi surface unless we
kill it. In the work of Luttinger \cite{Luttinger2}, Kohn and
Luttinger \cite{KL1}, Luttinger and Ward \cite{Ward} this process
ensures a nice perturbation series.

  \subsection{Quartic perturbations: the    RG  at Tree Level}
We now turn on the quartic interaction whose most general form is
\beq \delta S_4 = \frac{1}{2!2!}\int_{K\omega} \sib (4) \sib (3)
\psi (2) \psi (1) u(4, 3, 2 ,1) \label{s4} \eeq where \beqr \sib
(i) &=& \sib (K_i, \omega_i) \ \ etc., \\ \int_{K \omega} \! \!
&=&\! \!  \left[ \prod_{i=1}^{4}\int_{-
\pi}^{\pi}\frac{dK_i}{2\pi}\int_{-
\infty}^{\infty}\frac{d\omega_i}{2\pi}\right] \left[ 2\pi
\overline{\delta} (K_1 + K_2 - K_3 - K_4) 2\pi \delta (\omega_1 +
\omega_2 - \omega_3 - \omega_4 )\right] \label{measure1} \eeqr and
$\overline{\delta}$ enforces momentum conservation  mod $2\pi$, as
is appropriate to any lattice problem.  A process where lattice
momentum is violated in multiples of $2\pi$ is called an {\em
umklapp} process. The delta function containing frequencies
enforces time translation invariance.The coupling function $u$ is
antisymmetric under the exchange of its first or last two
arguments among themselves since that is true of the Grassmann
fields that it multiplies. {\em Thus the coupling $u$  has all the
symmetries of the full vertex function $\Gamma$ with four external
lines.}

Let us now return to the general interaction, Eqn.(\ref{s4} -
\ref{measure1}), and restrict the momenta to lie within $\Lambda$
of either Fermi point L or R. Using a notation where L (left Fermi
point) and R (right Fermi point) become  discrete a label $i = l \
or \ R$  and 1-4 label   the frequencies and momenta (measured
from the appropriate Fermi points).   Eqns.(\ref{s4} -
\ref{measure1} ) become

\beq \delta S_4 = \frac{1}{2!2!} \sum_{i_1 i_2 i_3 i_4= L,R}
\int_{K\omega}^{\Lambda} \sib_{i_4} (4) \sib_{i_3} (3) \psi_{i_2}
(2) \psi_{i_1} (1) u_{i_4 i_3 i_2 i_1}(4, 3, 2 ,1) \eeq where
\beqr \int_{K \omega} ^{\Lambda}  &=&  \left[ \int_{-
\Lambda}^{\Lambda}\frac{dk_1\cdots dk_4}{(2\pi    )^4}\int_{-
\infty}^{\infty}\frac{d\omega_1\cdots d\omega_4}{(2\pi )^4}\right]
\left[ 2\pi \delta (\omega_1 + \omega_2 - \omega_3 - \omega_4
)\right] \nonumber \\
                              & &\left[ 2\pi
\overline{\delta} ( \varepsilon_{i_1} (K_F + k_1 )
+\varepsilon_{i_2} (K_F +k_2) -  \varepsilon_{i_3} ( K_F + k_3) -
\varepsilon_{i_4} (K_F + k_4) )\right]\label{measure2}
 \eeqr
and \beq \varepsilon_i = \pm 1 \ \ for \ \ R\ , \ L . \eeq

Let us now implement the   RG    transformation    with this
interaction. This proceeds exactly as in $\phi^4$ theory. Let us
recall how it goes. If schematically \beq Z = \int d\phi_< d\phi_>
e^{-\phi^{2}_{<} - \phi^{2}_{>}} e^{-u(\phi_< + \phi_> )^4} \eeq
is the partition function and we are eliminating $\phi_>$, the
fast modes, the  effective $u$ for $\phi_<$ has two origins.
First, we have a term $-u \phi^{4}_{<} $ which is there to begin
with, called the {\em tree level} term. Next, there are terms
generated by the $\phi_>$ integration. These are computed in a
cumulant expansion and are given by Feynman diagrams whose
internal momenta lie in the range being eliminated. The loops that
contribute to the flow of $u$ begin at order $u^2$.

Let us first do the  order $u$ tree level calculation for the
renormalization of the quartic interaction. This gives us just
Eqn.(\ref{measure2}) with $\Lambda \rightarrow \Lambda /s$. If we
now rewrite this in terms of new momenta and fields, we get an
interaction with the same kinematical limits as before and we can
meaningfully read off the coefficient of the quartic-Fermi
operators as the new coupling function. We find \beq u'_{i_4 i_3
i_2 i_1}(k_i', \omega_i'  ) =u_{i_4 i_3 i_2 i_1}(k_i'/s, \omega_i'
/s ) \label{scaling} \eeq

The reader who carries out the intermediate manipulations will
notice  an important fact: $K_F$ never enters any of the $\delta$
functions: either all $K_F$'s cancel in the nonumklapp cases, or
get swallowed up in multiples of $2\pi$ (in inverse lattice units)
in the umklapp cases due to the periodicity of the
$\overline{\delta} $-function.  As a result the momentum $\delta$
functions are free of $K_F$ and  scale very nicely under the   RG
transformation: \beqr \overline{\delta}(k) &\rightarrow &
\overline{\delta}(k'/s)\\
                  &=& s \overline{\delta}(k')
\eeqr Turning now to Eqn.(\ref{scaling}), if we expand $u$ in a
Taylor series in its arguments and compare coefficients, we find
readily  that the constant term $u_0$  is marginal and the higher
coefficients are irrelevant. Thus $u$ depends only on its discrete
labels and we can limit the problem to just a few coupling
constants instead of the coupling function we started with.
Furthermore, all reduce to just one coupling: \beq u_0 = u_{LRLR}
= u_{RLRL}
      = -u_{RLLR}
        = -u_{LRRL} .
        \eeq
        Other couplings corresponding to $LL \rightarrow RR$ are wiped
out by the Pauli principle since they have no momentum dependence
and can't have the desired antisymmetry.

 The tree level analysis readily extends to couplings with six or more fields.
All these are irrelevant, even if we limit ourselves to constant
($\omega$ and k independent) couplings.

In summary, the RG tells us that the most general low energy
theory for spinless fermions in $d=1$ has, at tree level,  a
single marginal  coupling constant.

 To determine the ultimate fate of the coupling $u_0$, marginal at tree level,
we must turn to the one loop   RG  effects.

\subsection{  RG  at one loop: The Luttinger Liquid}

Let us  begin with the action with the quartic interaction and do
a mode elimination. Consult Figure 1 for details. To order $u$,
this leads to an induced quadratic term represented by the tadpole
graph in Figure 2. We set $\omega = k=0$ for the external legs
(since the dependence on these is irrelevant)  and have chosen
them to lie at L, the left Fermi point. The integral given by the
diagram produces a momentum independent term of the form $\delta
\mu \sib_L \psi_L$. But we began with no such term. Thus we do not
have a fixed point in this case. Instead we must begin with some
term $\delta \mu^{*}  \sib_L \psi_L$ such that upon
renormalization it reproduces itself. We find it by demanding that
\beq \delta \mu^* = s \left[  \delta \mu^* - u_0^* \int_{-
\infty}^{\infty}\frac{d\omega}{2\pi} \int_{\Lambda /s < |k| <
\Lambda} \frac{dk}{2\pi} e^{i\omega 0^+} \frac{1}{i\omega -
k}\right] \eeq where we have used the zeroth order propagator and
the fact that to this order any $u_0 = u_0^*$. The exponential
convergence factor is the one always introduced  to get the right
answer for, say, the ground state particle density using $<\sib
\psi >$. Doing the $\omega $ integral, we get \beqr \delta \mu^*
&= &s \left[ \delta \mu^* - u_0^* \int_{\Lambda /s < |k| <
\Lambda} \frac{dk}{2\pi} \theta (-k) \right] \\
              &=& s \left[\delta \mu^* - \frac{\Lambda u_0^*}{2\pi} ( 1-
1/s) \right] . \eeqr It is evident that the fixed point is given
by \beq \delta \mu^* = \frac{\Lambda u_0^*}{2\pi}.\label{deltamu}
\eeq

Alternatively, we could just as well begin with the following
relation for the renormalized coupling \beq \delta \mu^{'} = s
\left[  \delta \mu - u_0^* \int_{-
\infty}^{\infty}\frac{d\omega}{2\pi} \int_{\Lambda /s < |k| <
\Lambda} \frac{dk}{2\pi} e^{i\omega 0^+} \frac{1}{i\omega -
k}\right] \eeq which implies the flow \beq \frac{d\mu}{dt} =\mu
-\frac{u_{0}^{*}}{2\pi}\label{mueqn} \eeq assuming we  choose to
measure $\mu$ in units of $\Lambda$. The fixed point of this
equation reproduces Eqn.(\ref{deltamu}).

We can find $\delta \mu^*$ in yet another way with no reference to
the   RG . If we calculate the inverse propagator in the cut-off
theory to order u, we will find \beq G^{-1} = i\omega - k -
\frac{\Lambda u_0 }{2\pi} \eeq indicating that the Fermi point is
no longer given by $k = 0$. To reinstate the old $K_F$ as
interactions are turned on, we must move the chemical potential
away from zero and to the value $\delta \mu = \frac{\Lambda
u_0}{2\pi}$. Thus the correct action that gives us the desired
$K_F$, for this coupling, to this order, is then schematically
given by \beq S = \sib ( i \omega - k )\psi  + \frac{\Lambda
u_0}{2\pi}\sib \psi + \frac{u_{0}}{2!2!} \sib \sib \psi \psi .
\eeq An  RG    transformation  on this action would not generate
the tadpole graph contribution.

{\em A very important point which will appear again is this: we
must fine tune the chemical potential as a function of u, not to
maintain criticality (as one does in $\phi^4$ where the bare mass
is varied with the interaction to keep the system massless) but to
retain the same  particle density.} (To be precise, we are keeping
fixed $K_F$, the momentum at which the one-particle Greens
function has its singularity.  This amounts to keeping the density
fixed, following  Luttinger(1960).) If we kept $\mu$ at the old
value of zero, the system would flow away from the fixed point
with $K_F = \pi /2$, not to a state with a gap, but to another
gapless one with a smaller value of $K_F$. This simply corresponds
to the fact if the total energy of the lowest energy  particle
that can be added to the system, namely $\mu $, is to equal 0, the
kinetic energy at the Fermi surface must be slightly negative so
that the repulsive potential energy with the others in the sea
brings the total to zero.

Let us now turn our attention to the order $u_0^2$ graphs that
renormalize $u_0$. These are shown in Fig. 3. The increment in
$u_0$, hereafter simply called $u$, is given by the sum of the ZS
(zero-sound),  ZS' and BCS graphs. The analytical formula for the
increment in $u$ is \beqr du ( 4321 ) & =& \int u (6351) u(4526)
G(5) G(6) \delta (3+6-1-5) d5 d6 \nonumber \\
  & &      -\int u (6451) u(3526) G(5) G(6) \delta (6+4 - 1-5) d5 d6 \nonumber
\\
             & & - \frac{1}{2} \int u (6521) u (4365) G(5) G(6) \delta (5
+ 6 - 1  -2 ) d5 d6 \label{flow}
             \eeqr
where $1$ to $4$ stand for all the attributes of the (slow)
external lines, $5$ and $6$ stand for all the attributes of the
two (fast) internal lines: momenta (restricted to be within the
region being eliminated), and   frequencies;  G are the
propagators and the $\delta $ functions are for ensuring the
conservation of  momenta and frequencies and $\int d5d6$ stands
for sums and integrals over the attributes $5$ and $6$. (In the
figure the momenta $1$ to $6$ have been assigned some special
values (such as $5=K$ in Fig.3a) that are appropriate to the
problem at hand. The formula is very general as it stands and
describes other situations as well.) The couplings $ u$  are
functions of all these  attributes, with all the requisite
antisymmetry properties. (The order in which the legs are labeled
in $u$ is important due to all the minus signs. The above
equations have been written to hold with the indicated order of
arguments. In their present form they are ready to be used by a
reader who wants to include spin.)

This is the master formula we will invoke often. It holds even in
higher dimensions, if we suitably modify the integration region
for the momenta.

 Readers familiar
with Feynman diagrams may obtain this formula by drawing  all the
diagrams to this order in the usual Feynman graph expansion, but
allowing the loop momenta to range only over the modes being
eliminated. In the present case, these are given by the four thick
lines  labeled a,b,c and d in Fig. 1 where each  line  stands for
a region of width $d\Lambda $ located at the cut-off , i.e., a
distance $\Lambda$ from the Fermi points. The external momenta are
chosen to be $(4321) = (LRLR)$, at the Fermi surface. All the
external $k$'s and $\omega$'s are set equal to zero since the
marginal coupling $u$ has no dependence on these. This has two
consequences. First, the loop frequencies in the ZS and ZS' graphs
are equal, while those in the BCS graph are equal and opposite.
(The labels Zero sound and BCS describe the topology of these
graphs and not literally   these phenomena. ) Second, the momentum
transfers at the left vertex are $Q = K_1 - K_3 = 0$ in the ZS
graph, $Q' = K_1 - K_4 = \pi $ in the ZS' graph, while the total
momentum   in the BCS graph is $P = K_1 + K_2 = 0$. Therefore if
one  loop momentum $5=K$ lies in any of the four shells in Fig.1,
so does the other loop momentum $6$ which equals $K$, $K + \pi$ or
$-K$ in the ZS, ZS' and BCS graphs respectively. Thus we may
safely eliminate the momentum conserving $\delta $ function in
Eqn.(\ref{flow}) using $\int d6$. This fact, coupled with \beqr
 E(-K) &=& E(K) \\
E(K' =  K\pm \pi ) &=& -E(K) \eeqr leads to \beqr du(LRLR) \! \!
&=& \! \!  \int_{- \infty}^{\infty}\int_{d\Lambda}\frac{d\omega
dK}{4\pi^2} \frac{u(KRKR) u(LKLK)}{(i\omega - E(K))(i\omega -
E(K))} - \int_{-\infty}^{\infty}\int_{d\Lambda}\frac{d\omega
dK}{4\pi^2} \frac{u(K'LKR) u(RKLK')}{(i\omega - E(K))(i\omega +
E(K))}\nonumber \\
        &  &
-\frac{1}{2} \int_{-\infty}^{\infty}\int_{d\Lambda}\frac{d\omega
dK}{4\pi^2} \frac{u(-KKLR) u(LR-KK)}{(i\omega - E(K))(-i\omega -
E(K))} \label{oneloop} \\ &\equiv & ZS + ZS' + BCS
 \eeqr
 where $\int_{d\Lambda}$ means the momentum must lie in one of the four
slices in Fig.1.

 The reader is reminded once again that the names ZS, ZS' or BCS refer only
to the topologies of the graphs. To underscore this
point,especially for readers who have seen a similar integral in
zero sound calculations, we will now discuss the ZS graph. In the
present problem the loop momentum $K$ lies within a sliver
$d\Lambda$ of the cut-off. Both propagators have poles at the
point $\omega = -iE(k = \pm \Lambda)$. No matter which half-plane
this lies in, we can close the contour the other way and the
$\omega$ integral vanishes. This would be the case even if a small
external momentum transfer ($Q = K_3 - K_1 << \Lambda$) takes
place at the left vertex  since both poles would still be on the
same side. This is very different from what happens in zero sound
calculations where    the loop momenta roamed freely within the
cut-off, and in particular, go to the Fermi surface.  In that
case, the integral becomes very sensitive to how the external
momentum transfer $Q = K_3 - K_1 $ and  frequency transfer $\Omega
= \omega_3 - \omega_1$  are taken to zero since  any nonzero $Q$,
however small, will split the poles and make them lie on different
half planes for $k < Q $ and the integral will be nonzero.  It is
readily seen that
 \beq
 \int_{-\infty}^{\infty}\int_{-\Lambda}^{\Lambda} \frac{d\omega
dk}{4\pi^2} \frac{1}{(i\omega - k)(i\omega  -k - Q +i\Omega ) } =
\int_{- \Lambda}^{\Lambda} \frac{dk}{2\pi} \frac{i}{\Omega +iq}
(\theta (k) - \theta (k+q))
 \eeq
 where the step function $\theta (k)$ is is simply related to the Fermi function:
$f(k) = 1- \theta (k)$.    If we keep $\Omega \ne 0$ and send $Q$
to zero we get zero. On the other hand of we set $\Omega = 0$ and
let $Q$ approach zero we get (minus) the derivative of the (Fermi)
$\theta $ function, i.e., a $\delta$-function at the Fermi surface
. Thus reader used to zero-sound physics should not be disturbed
by the fact that the ZS graph makes  no contribution  since the
connotation here is different.

Now  for the ZS' graph, Fig.3b, Eqn.(\ref{oneloop}). We see that
$K$ must lie near $L$ since $1=R$ and there is no RR scattering.
As far as the coupling at the left vertex is concerned, we may set
$K =  L$ since the marginal coupling has  no $k$ dependence. Thus
$K + \pi = R$ and the vertex becomes $u(RLLR) = -u$. So does  the
coupling at the other vertex. Doing the $\omega$ integral (which
is now nonzero since the poles are always on opposite half-planes)
we obtain, upon using the fact that there are two shells (a and b
in Fig.1) near L and that $|E(K)| = |k| = |\Lambda |$, \beqr ZS'
&=& u^2 \int_{d\Lambda \in L} \frac{dK}{4\pi |E(K)|} \nonumber \\
 &=& \frac{u^2}{2\pi} \frac{d|\Lambda |}{\Lambda}
 \label{ZS'}
 \eeqr
The reader may wish to check that the ZS' graph will make the same
contribution to the $\beta$-function in the field theory approach.

 The BCS graph (Eqn.(\ref{oneloop}), Fig.3c) gives a nonzero contribution
since the propagators have opposite frequencies, opposite momenta,
but equal energies due to time-reversal invariance $E(K)= E(-K)$.
We notice that the factor of $\frac{1}{2}$ is offset by the fact
that K can now lie in any of the four regions a,b,c, or d. We
obtain
 a contribution of the same magnitude but opposite sign as ZS' so that
 \beqr
 du &=& (\frac{u^2}{2\pi} - \frac{u^2}{2\pi} ) \underbrace{\frac{d|\Lambda
|}{\Lambda}}_{dt} \\
  \frac{du}{dt}      &=& \beta (u) = 0.
  \eeqr

  Thus we find that $u$ is {\em still} marginal.  The flow to one loop
  for $\mu$ and $u$ is
  \beqr
  \frac{d\mu}{dt}& =& \mu - \frac{u}{2\pi}\\
  \frac{du}{dt}&=& 0.
  \eeqr
  There is a line of fixed points:
  \beqr
  \mu & =&  \frac{u^{*}}{2\pi}\\
      u^{*} & & arbitrary.
      \eeqr
  Notice that $\beta$ vanishes due to a  cancellation between two diagrams,
each of which by itself would have led to the CDW or BCS
instability.  When one does a mean-field calculation for CDW, one
focuses on just the ZS' diagram and ignores the BCS diagram. This
amounts to taking
  \beq
  \frac{du}{dt} = \frac{u^2}{2\pi}
  \eeq
  which, if correct, would imply that any positive $u$
  grows under renormalization. If this growth continues we expect a CDW.
  On the other hand,  if just the BCS diagram is
  kept we will  conclude a run-off for negative couplings leading to a state
with $<\psi_R \psi_L > \neq 0$.

            What the $\beta$ function does  is to treat these
            competing instabilities simultaneously  and predict  a
            scale-invariant theory.

             Is this the correct prediction for the spinless model?
             If we consider a nearest-neighbor interaction
             of strength $u$, the
exact solution of Yang and Yang (1976)\cite{Yang} tells us there
is no no gap till $u$ is of order unity. If the RG  analysis were
extended to higher loops we would keep getting
            $\beta = 0$ to all orders. This follows
             from the
            the Ward identity in the cut-off continuum model
            (Di Castro and Metzner \cite{metzner}) which reflects the fact that in
this  model, the number of fermions of type $L$ and $R$ are
separately conserved. The vanishing beta function also agrees with
the original finding of Solyom \cite{Solyom}.

This model also coincides with the massless Thirring model, which
is a Lorentz invariant  theory with a  current-current
interaction. The reason is that using $L$ or $R$ fields and  their
adjoints, there is just one possible quartic interaction. As for
Lorentz invariance, it was assured when we linearized the spectrum
near the Fermi points.

This scale-invariant system is called the Luttinger liquid. The
system has a Fermi surface at which the occupation number has a
kink in slope, but no jump in value. The Fermionic Green's
functions fall with anomalous powers that vary with the coupling
$u$. The quasi-particle is totally gone, no matter how small the
interaction. Under  bosonization,  the model maps into a free
bosons. The complicated fermionic behavior is encoded in the
expressions for  the fermions operators in the bosonic language.
If spin is included, we get two bosons, moving at different
velocities, a phenomenon called spin-charge separation.

 How do we ever reproduce
            the eventual charge density wave instability
             known to exist in the exact
            solution of the model with nearest neighbor interactions?
             The answer is as follows. As we move along the
            line of fixed points, labeled by $u$,  the dimension of
            various operators will change from the free-field values.
Ultimately the umklapp coupling , ($RR \leftrightarrow LL$) ,
which was suppressed by a factor $(k_1 - k_2 )(k_3 - k_4)$ , will
become marginal and then relevant, as shown by
Haldane\cite{Haldane}.  If we were not at half-filling such a term
would  be ruled out by momentum conservation and the scale
invariant  Luttinger liquid would persist for all $u$. While this
liquid provides us with an example of where the   RG does better
than mean-filed theory, it is rather special and seems to occur in
$d=1$ systems where the two Fermi points satisfy the conditions
for {\em both} CDW and BCS instabilities. In higher dimensions one
finds that any instability due to a divergent susceptibility is
never precisely cancelled by another.

\section{Higher Dimensions}
The extension of these methods to higher dimensions is discussed
in the RMP article and this discussion will be limited to the part
that makes contact with Luttinger's work.

In $d=2$  dimensions, I considered Fermi surfaces of arbitrary
shape. In the circular case, to which I limit myself here, I found
that if one took an annulus of thickness $2\Lambda$ concentric
with the Fermi circle, and let $\Lambda \to 0$, RG yielded two
marginal {\em coupling functions}, $F(\theta $) and $V(\theta )$,
see Figure 4. (As in $d=1$, the chemical potential had to be fine
tuned to keep $K_F$ fixed.) The function $F$  described forward
scattering and was shown to be
 Landau's $F$ function. It remained marginal to all orders in the
 loop expansion. (I pointed out that $\Lambda /K_F$ played the role
 of the small parameter $1/N$ which made such  statements
 possible.)
 The function $V(\theta )$, which described scattering
 between Cooper pairs,  evolved as follows:
 \beq
 {dV_m\over dt} = -c_m V^{2}_{m}
 \eeq
 where $V_m$ was the $m$-the Fourier coefficient (angular momentum
 $m$ channel of Cooper pairs). This
 meant that any positive $V_m$ flowed to zero logarithmically (an
 old result of Morel and Anderson \cite{AM} ) while {\em any
 attractive $V_m$ grew in strength leading to the BCS
 instability.}
 In other words,  the Fermi liquid had an infinite number of unstable
 directions, and attraction in any angular momentum channel between
 Cooper pairs spelled its end.

 We relate  the last statement with the result of
 Kohn-Luttinger\cite{KL} that any Fermi system will end up
 superconducting as follows. The couplings $V_m$ are are not the
 bare couplings, say of a Hubbard or continuum model. They are the
 result of integrating out all modes outside the cut-off,
say, using  in perturbation theory.
 The Kohn-Luttinger
result amounts to the statement that if one does this, some $V_m$
or other will surely be negative. To see this, one need look at
only the diagrams considered by Kohn-Luttinger. The fact that the
loop momenta lie outside the cut-off in the RG approach do not
change the conclusion that an attractive coupling will be
generated from repulsive ones, as shown in my review.
\section{Conclusions}
I have focused here on the results of Luttinger's work that arose
naturally when I used the RG as a tool to study many-body
fermionic systems. I discussed a special $d=1$ system in which
superconductivity and CDW instability duke it out and neither
wins, leaving behind a scale invariant system, the Luttinger
liquid. It is not possible to get such a system in $d>1$ without
invoking more singular interactions than the coulomb or strong
coupling. I showed that in order to zero-in on the low energy
sector for femions, namely the Fermi surface, one needs to know
where it lies and here the results of Luttinger, Kohn and Ward
reappear. Finally I showed that the Kohn-Luttinger result, of the
inevitable superconducting instability, appears here as the
statement that by the time we eliminate modes outside the cut-off
and get to the physics near the Fermi surface, some Cooper
coupling will surely become negative,  at which point  the RG
takes over and predicts it will grow.

I thank the organizers of this issue for giving me a chance to pay
tribute to the work of a physicist whom  I greatly admired and to
the National science Foundation for grant DMR98-00626.

\section {Captions}
{\bf Figure 1} The figure shows the regions
   of momentum space being integrated out in the $d=1$ spinless fermion
   problem. The thick lines stand for the slices of width
   $d\Lambda$. They lie a distance $\Lambda$
   from the Fermi points L and R. In the ZS graph which has zero momentum transfer, both lines lie on the same
   slice  and the $\omega$-integral gives zero. In the ZS' graph,  the
   momentum transfer  $\pi$, connects  $a$ and  $c$ (which have opposite energies)
   and $b$ and $d$ similarly related. In the BCS diagram the loop momenta are
   equal and opposite and correspond to $a,d$ or $b,c$.
   \\

{\bf Figure 2} The tadpole graph which renormalizes the fermion at
one loop. It has no dependence on $k$, the deviation of the
external momentum from $K_F$  or $\omega$. We have used this
freedom to set both these to zero on the external legs. The effect
of this graph may be neutralized by a counter-term corresponding
to a change in chemical potential. One may do this if one wants to
preserve $K_F$.\\

{\bf Figure 3} The one loop graphs for $\beta (u)$ for the
spinless
 fermions in $d=1$.  The loop  momenta  lie in the shell of width
 $d\Lambda$ being eliminated. The external frequencies being all
 zero, the loop frequencies are equal for ZS (Fig.3a) and ZS' (Fig.3b) graphs and
 equal and opposite for the BCS graph (Fig.3c).  The ZS graph does not
 contribute since both loop momenta are equal (the momentum
 transfer $Q$  at the left vertex is $0$) and  lie a distance
 $\Lambda$ above or below  the Fermi surface and the $\omega$
 integration vanishes  when the poles  lie on the same half-plane.
 The ZS' graph has momentum transfer  $\pi$ at the left and right
 ends. This changes the sign of the  energy of the line entering
 left the vertex. The $\omega$ integral is nonzero, the poles being on
 opposite half-planes. The BCS graph (c) also survives since the
 momenta
 loop momenta  are equal and opposite (since the incoming momentum
 is zero) and this again makes the poles go to the opposite half-planes
 because the lines have {\em opposite }frequencies. The labels $1\ldots 6$
 refer to the master Equation.(34).\\

{\bf Figure 4} (a): The quartic  coupling. A label like $1$ stands
for three things: an angle $\theta_1$ on the Fermi surface, a
frequency $\omega_1$
 and a momentum $K-K_F\equiv k$, both equal to $0$ since the
 dependence on these two is irrelevant. (b) The low energy
region that survives under RG in $d=2$. The bandwidth $2\lambda$
has become as as small as the thickness of the circular line. Note
that if the two incoming momenta lie on this circle, the outgoing
momenta must equal them: $\theta_1=\theta_3$ and
$\theta_2=\theta_4$ (or  the  exchanged version). The angle
$\theta_1-\theta_2=\theta $ is the argument of Landau's $F $
function. An  exception arises  if $\theta_1=-\theta_2$, in which
case $\theta_3=-\theta_4$ and the angle between these two opposing
lines is the argument of the Copper amplitude $V$.
\end{document}